\documentclass{iaus}
\usepackage{graphicx}

\title[The origin of lopsidedness in galaxies] 
{The origin of lopsidedness in galaxies}

\author[M. Mapelli, B. Moore \& J. Bland-Hawthorn]   
{M. Mapelli$^1$, B. Moore$^1$
 \and J. Bland-Hawthorn$^2$}

\affiliation{$^1$Istitute for theoretical physics, University of Zurich,\\ Winterthurerstrasse 190, CH 8057, Z\"urich, Switzerland \\email: {\tt mapelli@physik.unizh.ch}\\[\affilskip]
$^2$Institute of Astronomy, School of Physics, \\ University of Sydney,  Australia}

\pubyear{2008}
\volume{xxx}  
\pagerange{xxx}
\jname{The Galaxy Disk in Cosmological Context}
\editors{J. Andersen (Chief Editor), J. Bland-Hawthorn \&{} B. Nordstr\"om, eds.}
\begin{document}

\maketitle

\begin{abstract}
It has long been known that a large fraction of disc galaxies are lopsided. We simulate three different mechanisms that can induce lopsidedness: flyby interactions, gas accretion from cosmological filaments and ram pressure from the intergalactic medium. Comparing the morphologies, HI spectrum and $m=1$ Fourier components, we find that all of these mechanisms can induce lopsidedness in the gaseous component of disc galaxies. In particular, we estimate that flybys can contribute to $\sim{}20$ per cent of lopsided galaxies. We compare our simulations with the observations of NGC~891, a lopsided, edge-on galaxy with a nearby companion (UGC~1807). We find that the main properties of NGC~891 favour a flyby event for
the origin of lopsidedness in this galaxy.
\end{abstract}

\firstsection 
\section{Introduction}
A high fraction of disc galaxies are lopsided, i.e. their gas and/or stellar component extend further out on one side of the galaxy than on the other (\cite[Baldwin, Lynden-Bell \&{} Sancisi 1980]{Baldwin+80}; \cite[Richter \&{} Sancisi 1994]{richter94}; \cite[Rix \& Zaritsky 1995]{rix95}; \cite[Zaritsky \& Rix 1997]{zaritsky97}; \cite[Swaters et al. 1999]{swaters+99}; \cite[Bournaud et al. 2005]{bournaud+05}; see \cite[Sancisi et al. 2008]{sancisi+08} for a review). 

\cite[Richter \&{} Sancisi (1994)]{richter94} show that the lopsidedness of a galaxy can be inferred from asymmetries in its global HI profile, and estimate, from the analysis of 1700 HI spectra, that $\gtrsim{}50$ per cent of disc galaxies are lopsided in the gaseous component.
The kinematics of the gas is often affected by lopsidedness: \cite[Swaters et al. (1999)]{swaters+99} find that the rotation curve of lopsided galaxies is rising more steeply on one side than on the other.
\cite[Rix \& Zaritsky (1995)]{rix95} and \cite[Zaritsky \& Rix 1997]{zaritsky97}, using near-infrared photometry of nearly face-on spiral galaxies, show that even the stellar component is lopsided in $\sim{}30$ per cent of their sample. 

The hypothesis that lopsidedness is due to galaxy interactions has been long discussed.
Based on optical images, \cite[Odewahn (1994)]{odewahn+94} finds that 71 of 75 lopsided Magellanic spirals have a nearby companion. However, \cite[Wilcots \& Prescott (2004)]{wilcots04} obtain HI data of 13 galaxies from \cite[Odewahn (1994)]{odewahn+94} and show that only four of them have HI-detected neighbours.
Thus, either lopsidedness is not related to galaxy interactions, or the asymmetries produced by these interactions are long-lived (surviving for $\gtrsim{}2$ orbital times after the encounter) and the lopsidedness persists even when the companion is quite far-off.

From the theoretical point of view, the N-body simulations by \cite[Walker, Mihos \& Hernquist (1996)]{walker+96} suggest that minor mergers can induce lopsidedness over a long timescale ($\gtrsim{}$ 1 Gyr). However,  \cite[Bournaud et al. (2005)]{bournaud+05} indicate that 
the lopsidedness produced by minor mergers disappears when the companion is completely disrupted. Since most of observed lopsided galaxies are not undergoing mergers, the minor-merger scenario does not seem viable. 
\cite[Bournaud et al. (2005)]{bournaud+05} indicate that the most likely mechanism to produce lopsidedness is the accretion of gas from cosmological filaments.
Alternative models suggest that baryonic lopsidedness can be induced by a lopsided dark matter halo (\cite[Jog 1997]{jog97}) or by the fact that the disc is off-centre with respect to the dark matter halo (\cite[Levine \& Sparke 1998]{levine98}). 
 
Here we address the problem of the origin of lopsidedness by means of $N-$body/smoothed particle hydrodynamics (SPH) simulations. In particular, we re-analyze in more detail the hypothesis of gas accretion, already proposed by \cite[Bournaud et al. (2005)]{bournaud+05}, and we consider two new possible scenarios: the role of flyby interactions with smaller companions and that of ram pressure from the intergalactic medium (IGM).


\begin{figure}[b]
\begin{center}
 \includegraphics[width=6.2in]{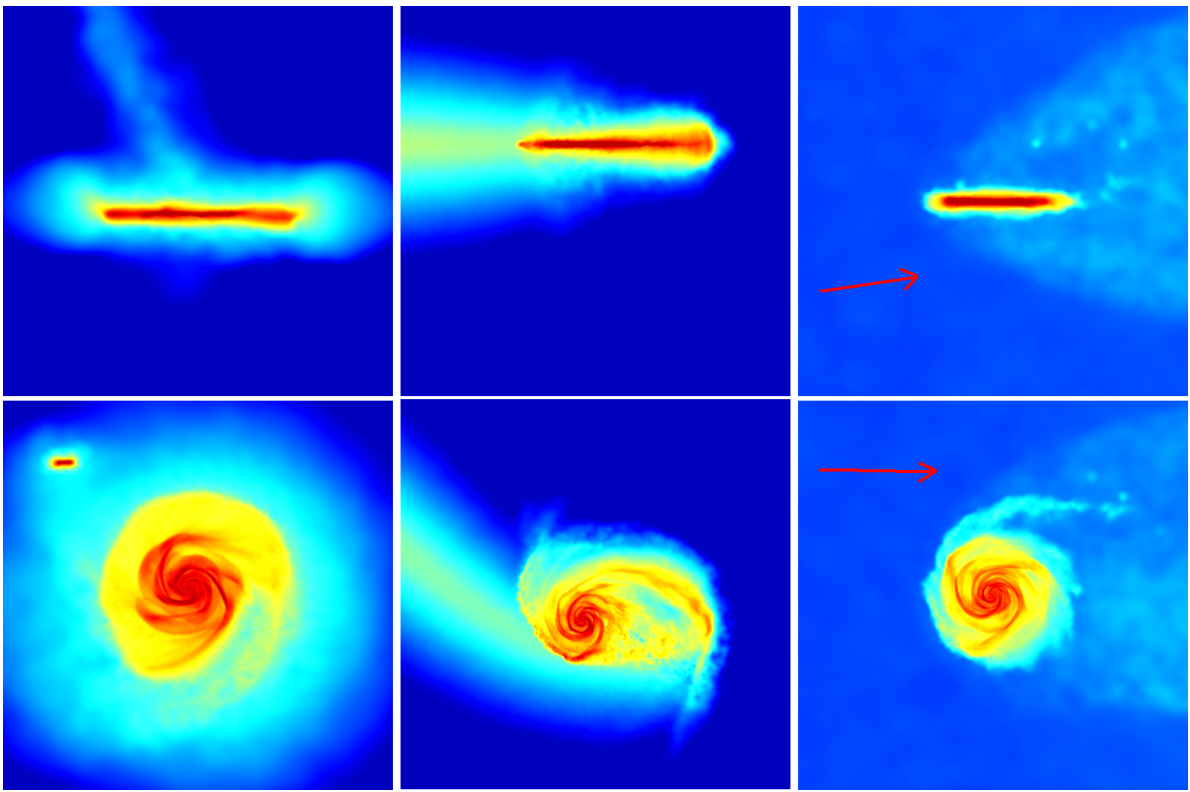} 
 \caption{Projected density of gas in the simulations. Top (Bottom) panels: the target galaxy is seen edge-on (face-on). From left to right: flyby, gas accretion and ram pressure scenario. The left-hand panels measure 80 kpc per edge. The central and the right-hand panel measure 100 kpc per edge. The time elapsed from the beginning of the simulation is $0.64$, 0.72 and 1.0 Gyr in the flyby, gas accretion and ram pressure case, respectively. The density goes from $2.23\times{}10^{-3}$ to $2.23\times{}10^1\,{}M_\odot{}$~pc$^{-2}$, from $5.60\times{}10^{-2}$ to $2.23\times{}10^1\,{}M_\odot{}$~pc$^{-2}$ and from $7.05\times{}10^{-2}$ to $2.23\times{}10^1\,{}M_\odot{}$~pc$^{-2}$, in logarithmic scale, in the flyby, gas accretion and ram pressure case, respectively.}
   \label{fig1}
\end{center}
\end{figure}

\section{Methods}
We consider three different processes: flybys (i.e. interactions between a target and an intruder galaxy which do not imply merger), gas accretion from cosmological filaments (\cite[Keres et al. 2005]{keres+05}) and ram pressure from low-density IGM. For each of these cases we make a large number of simulations with good spatial and mass resolution (the softening length and the mass of each particle are 0.1 kpc and $2.23\times{}10^4\,{}M_\odot{}$ for gaseous and stellar particles, and 0.2 kpc and $2.23\times{}10^5\,{}M_\odot{}$ for dark matter particles, respectively). The $N-$body/SPH code used is GASOLINE (\cite[Wadsley, Stadel \& Quinn 2004]{wadsley+04}). The details of the simulations have been described in \cite[Mapelli et al. (2008a)]{mapelli+08a}. Here we provide only the fundamental characteristics.

{\underline{\it Flybys.}}
We make different runs, with different mass ratios between the intruder and the target (1:20 to 1:2), with different masses of the intruder and of the target, with different impact parameter and relative velocity.

{\underline{\it Gas accretion.}}
The simulated galaxy accretes a cylinder of uniform gas with relative velocity of 100 km s$^{-1}$ and with mass accretion $\dot{M}=2-6\,{}M_\odot{}$. Different runs have been made with different inclinations of the gas cylinder with respect to the plane of the galaxy. This model, although very simplified, is supported by the gaseous filaments observed in cosmological simulations (\cite[Keres et al. 2005]{keres+05}).

{\underline{\it Ram pressure.}}
The determination of the density of IGM in the field and in poor groups is currently an open issue. Recent observations (\cite[Pisano et al. 2004]{pisano+04}; \cite[Freeland, Cardoso \& Wilcots 2008]{freeland+08}) suggest that the IGM density in groups is $\gtrsim{}10^{-5}$ cm$^{-3}$. At such densities, relative velocities $\gtrsim{}100$ km s$^{-1}$ should lead to ram-pressure stripping (\cite[Gunn \& Gott 1972]{gunn72}). Thus, we simulate a galaxy moving with a velocity of $200$ km s$^{-1}$ in a uniform IGM with density $5\times{}10^{-5}$ cm$^{-3}$.

\begin{figure}[b]
\begin{center}
 \includegraphics[width=6.2in]{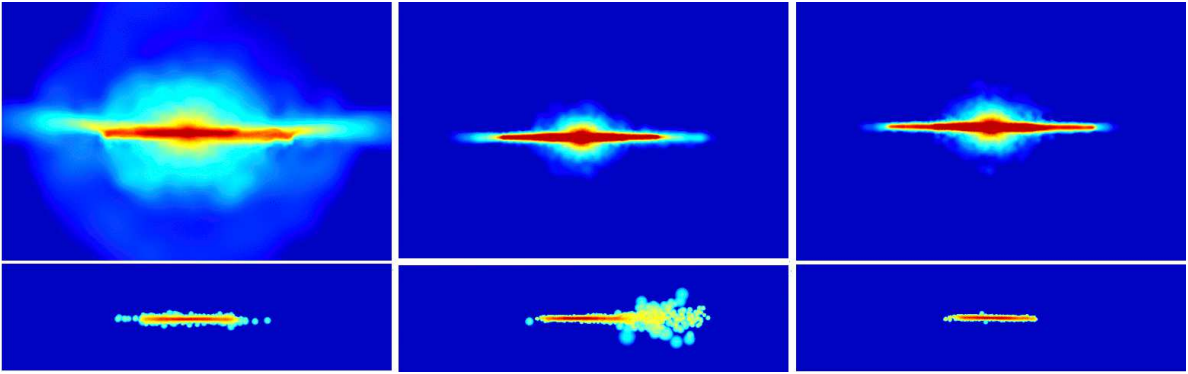} 
 \caption{Projected density of stars in the simulations. Top panels: all stars. Bottom panels: young stars ($\lesssim{}100$ Myr). From left to right: flyby, gas accretion and ram pressure scenario. The left-hand panels measure 80 kpc along the long side. The central and the right-hand panel measure 100 kpc along the long side. The time elapsed from the beginning of the simulation is $0.64$, 0.72 and 1.0 Gyr in the flyby, gas accretion and ram pressure case, respectively. Top (bottom) panels: the density goes from $2.23\times{}10^{-3}$ to $2.23\times{}10^2\,{}M_\odot{}$~pc$^{-2}$ (from $7.05\times{}10^{-6}$ to $2.23\times{}10^2\,{}M_\odot{}$~pc$^{-2}$), from $7.05\times{}10^{-2}$ to $2.23\times{}10^1\,{}M_\odot{}$~pc$^{-2}$ (from $7.05\times{}10^{-6}$ to $2.23\times{}10^1\,{}M_\odot{}$~pc$^{-2}$) and from $5.60\times{}10^{-2}$ to $2.23\times{}10^1\,{}M_\odot{}$~pc$^{-2}$ (from $7.05\times{}10^{-6}$ to $2.23\times{}10^1\,{}M_\odot{}$~pc$^{-2}$),  in logarithmic scale, in the flyby, gas accretion and ram pressure case, respectively.}
   \label{fig2}
\end{center}
\end{figure}

\section{Results}
The main result of our simulations is that all these mechanisms (flybys, gas accretion and ram pressure) produce lopsidedness in the gaseous component of the simulated galaxy. This can be seen qualitatively by the density maps shown in  Figure 1. Such result is important, as it indicates that lopsidedness is not due to a single process, but may originate from different mechanisms.
The total stellar component of the galaxy is significantly lopsided only in the case of flybys (upper panels of Figure 2). Interestingly, the young stellar component  (i.e. the stellar particles with an age $\lesssim{}100$ Myr) is strongly lopsided also in the gas accretion scenario (bottom panel of Figure 2). This indicates that, in the gas accretion scenario, the lopsidedness of the stars may be induced by star formation in a lopsided gaseous component.
A more quantitative estimate of lopsidedness is given by the normalized strength of the Fourier component $m=1$ (A1) as a function of radius, shown in the left-hand panel of Figure 3. The Fourier analysis confirms that all the three mechanisms induce lopsidedness in the gaseous component. The flyby scenario produces lopsidedness both in the old and in the young stars, the gas accretion scenario only in the young stars, whereas the ram-pressure model does not induce lopsidedness in the stars.
\begin{figure}[b]
\begin{center}
 \includegraphics[width=2.5in]{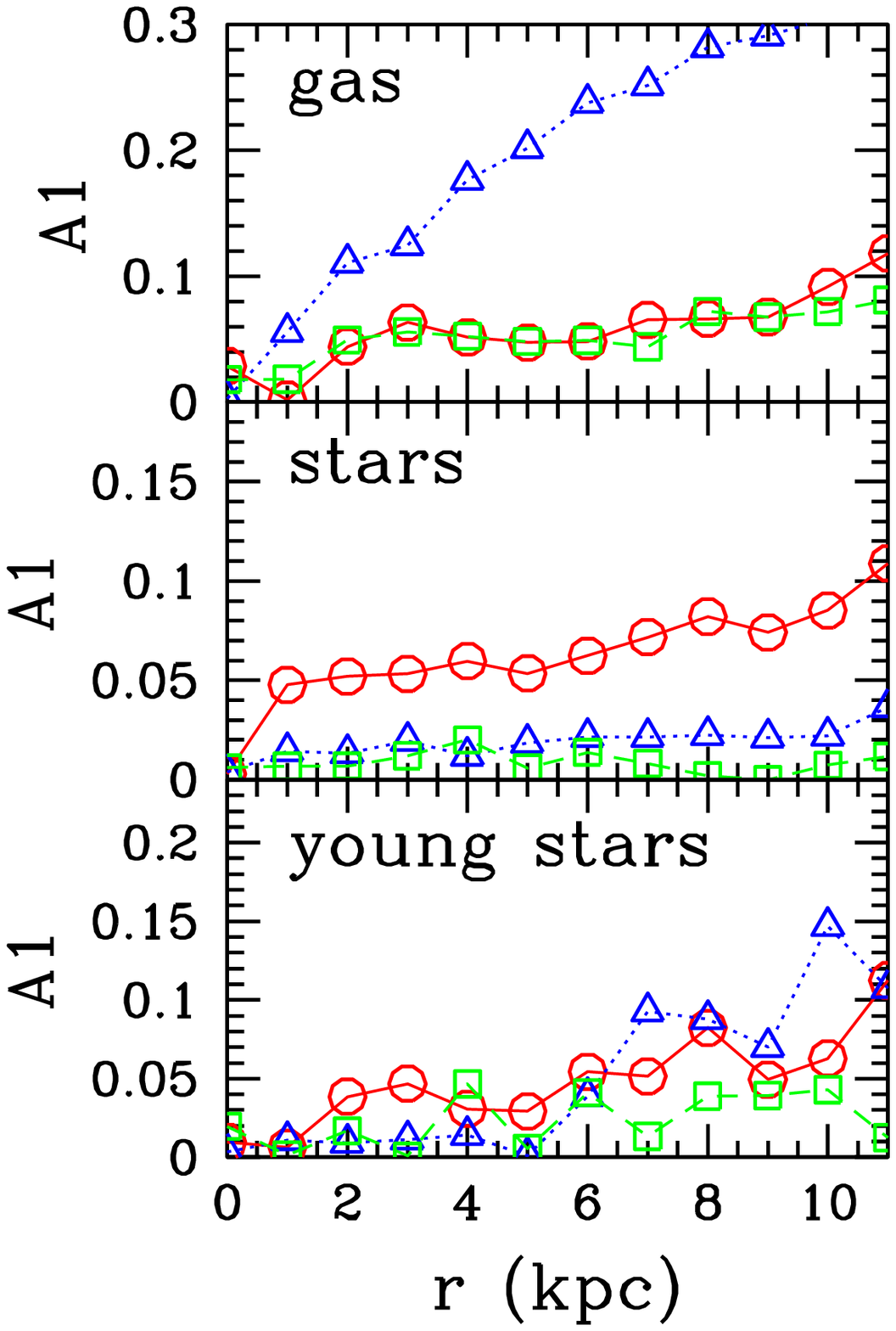} 
 \includegraphics[width=2.5in]{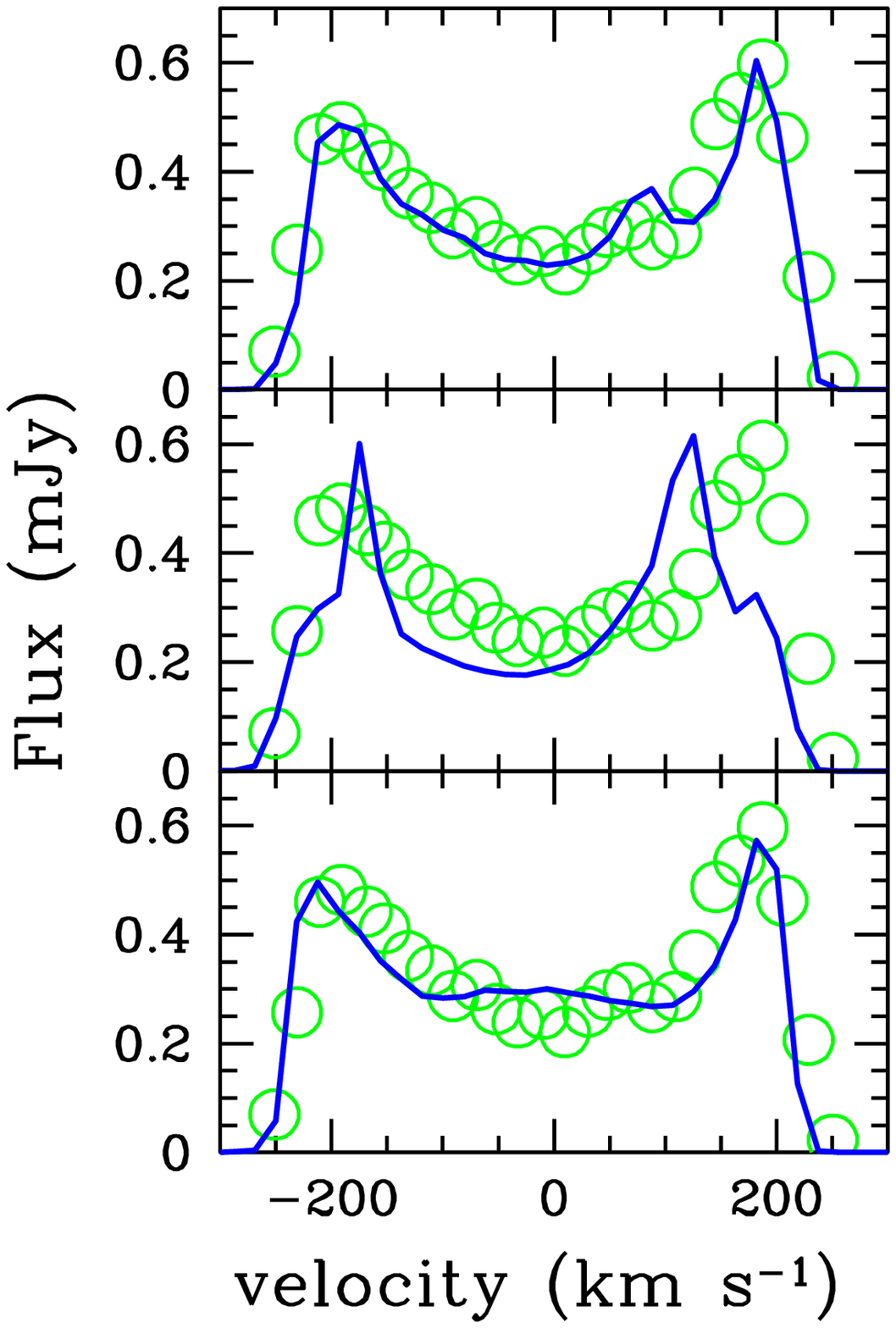} 
 \caption{Left-hand panel: Normalized strength of the Fourier component $m=1$ as a function of radius, for  gas (top panel), global stellar content (central panel) and young stellar population (bottom panel). Open circles connected by red solid line: flyby scenario; open triangles connected by blue dotted line: gas accretion; open squares connected by green dashed line: ram pressure. Right-hand panel: HI spectra of the simulated galaxies compared with the observations of NGC~891. Open green circles: data from \cite[Richter \& Sancisi (1994)]{richter94}. Solid blue line: simulations. Top panel: flyby interaction; central panel: gas accretion; bottom panel: ram pressure.}
   \label{fig3}
\end{center}
\end{figure}

We also analyze the kinematic properties (HI spectrum and rotation curve) of the simulated galaxies. The HI spectrum is shown in the right-hand panel of Figure~3. The simulations (blue line) are compared with the data of the lopsided galaxy NGC~891 (green circles, from \cite[Richter \&{} Sancisi 1994]{richter94}). The main feature of lopsidedness in the HI spectra is the asymmetry of the two peaks of the spectrum. Our simulated galaxies show only a moderate lopsidedness in the HI spectra. In particular, the flyby and the ram-pressure case agree very well with the data of NGC~891.

\begin{figure}[b]
\begin{center}
 \includegraphics[width=3.2in]{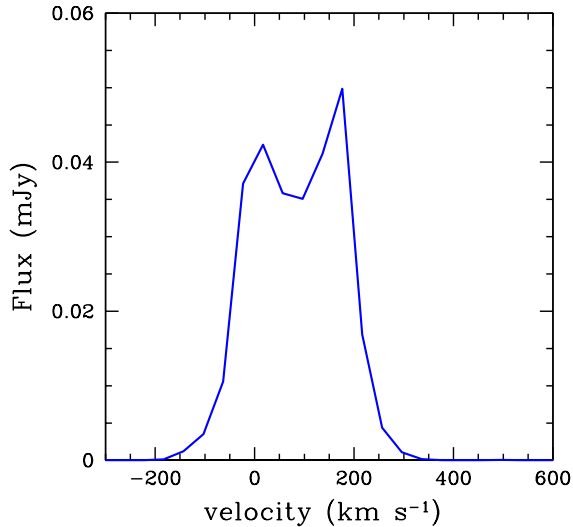} 
 \caption{HI spectrum of the simulated intruder galaxy (i.e. the small companion) in the flyby scenario.}
   \label{fig4}
\end{center}
\end{figure}
\subsection{Flyby rate}
We have shown that flybys can induce lopsidedness in galaxies, both in the gaseous and in the stellar population. We now derive which fraction of disc galaxies may be lopsided due to flybys (an analogous estimate is currently hard to do for gas accretion and ram pressure, given the insufficient observational constraints). The density of galaxies which are lopsided as a consequence of flybys may be derived from the density of collisional ring galaxies (CRGs, see \cite[Mapelli et al. 2008b]{mapelli+08b} and references therein) in the local Universe ($n_{\rm CRG}=5.4\times{}10^{-6}\,{}h^3\,{}$Mpc$^{-3}$, where $h$ is the Hubble parameter, \cite[Few \& Madore 1986]{few86}). In fact, the galaxy interactions which induce lopsidedness are similar to the ones which produce the CRGs, although with some crucial difference (e.g. the range of impact parameters, the inclination angles, the lifetimes). Taking into account this analogy, the density of galaxies which are lopsided due to a flyby ($n_{\rm fly}$) can be derived as:
\begin{equation}\label{eq:eqring}
n_{\rm fly}\sim{}n_{\rm CRG}\,{}\frac{b_{\rm max}^2-b_{\rm CRG}^2}{b_{\rm CRG}^2}\,{}\frac{1}{\left(1-\cos{45^\circ{}}\right)}\,{}\frac{t_{\rm lop}}{t_{\rm CRG}}, 
\end{equation}
where $b_{\rm max}\sim{}4\,{}R_d$ and $b_{\rm CRG}\lesssim{}0.36\,{}R_d$ are the maximum impact parameter to induce lopsidedness and to form a CRG, respectively ($R_d$ is the disc scale-length), whereas $t_{\rm lop}$ and $t_{\rm CRG}$ are the lifetime of lopsidedness and of a CRG, respectively. The term $\left(1-\cos{45^\circ{}}\right)$ accounts for the fact that only interactions with inclination angle $\le{}45^\circ{}$ produce CRGs (\cite[Few \& Madore 1986]{few86}). Assuming $t_{\rm lop}\sim{}1$ Gyr and $t_{\rm CRG}\sim{}0.5$ Gyr, as suggested by simulations, we obtain $n_{\rm fly}\sim{}4.5\times{}10^{-3}\,{}h^3$ Mpc$^{-3}$. Since the density of relatively bright disc galaxies (i.e. with absolute magnitude $\lesssim{}-19$ mag) is $n_{\rm disc}\sim{}4.9\times{}10^{-2}\,{}h^3$ Mpc$^{-3}$ (\cite[Few \& Madore 1986]{few86}), this suggests that only one galaxy every $\sim{}11$ spiral galaxies can be lopsided as a consequence of flybys. As lopsidedness of the gas component is observed in $\sim{}50$ per cent of disc galaxies (\cite[Richter \& Sancisi 1994]{richter94}), flybys can reasonably contribute to $\sim{}20$ per cent of lopsided galaxies.

\subsection{Comparison with NGC~891}
Finally, we compare the properties of the simulated galaxy with those of one of the most studied lopsided galaxies, NGC~891 (\cite[Oosterloo, Fraternali \&{} Sancisi 2007]{oosterloo07} and references therein). This galaxy is mildly lopsided (right-hand panel of Figure 3) and has some interesting features. In particular, NGC~891 has a conspicuous HI halo ($\sim{}30$ per cent of the total observed HI). Furthermore, the observations show a gaseous filament extending up to $\sim{}20$ kpc vertically from the disc and located at $\sim{}10$ kpc from the centre of the galaxy. Finally, NGC~891 has a smaller, gas-rich companion, UGC~1807, located at a projected distance of $\sim{}80$ kpc, in the direction of the above mentioned gaseous filament.

The simulations reproduce quite well the lopsidedness of NGC~891\footnote{The simulations presented in \cite[Mapelli et al. 2008a]{mapelli+08a} assume a dark matter halo of $1.4\times{}10^{11}\,{}M_\odot{}$. Such mass is likely lower than the mass of the dark matter halo of NGC~891. However, check simulations with a dark matter mass of $3.5\times{}10^{11}\,{}M_\odot{}$ and of $5.0\times{}10^{11}\,{}M_\odot{}$ show that the main conclusions are unchanged, when changing the mass by a factor of $\sim{}2-3$.}. In particular, the flyby and the ram pressure scenarios match the HI spectrum (right-hand panel of Figure 3). However, the gas accretion and the ram pressure simulations cannot reproduce the observed filament. Only the flyby scenario produces a sort of gaseous filament between the target and the intruder (left-hand panel of Figure 1). This filament is similar to the one observed in NGC~891, although there are some differences (e.g. the simulated filament extends with almost uniform density between the target and the intruder, whereas the observed filament is denser in the region which is close to NGC~891). Thus, our simulations suggest that the lopsidedness of NGC~891 may be due to flybys. However, further comparisons with data are required. For example, in Figure~4 the HI spectrum of the simulated intruder galaxy is shown. This spectrum appears mildly lopsided, similarly to that of the target galaxy. It would be interesting to look at the spectrum of UGC~1807 (and, in general, of the companions of other lopsided galaxies) and to see if they appear lopsided, too.
\begin{acknowledgements}
We thank F.~Bournaud, T.~Oosterloo, R.~Sancisi and B.~Gibson for useful discussions.
\end{acknowledgements}

\end{document}